\documentclass[twocolumn,showpacs,aps,prb,preprintnumbers,amsmath,amssymb,superscriptaddress]{revtex4}
\usepackage{graphicx}
\usepackage{amssymb}

\newcommand{\R}{\mathbf{r}}

\newcommand{\UP}{n_{\uparrow}}
\newcommand{\DN}{n_{\downarrow}}
\newcommand{\TUP}{\tau_{\uparrow}}
\newcommand{\TDN}{\tau_{\downarrow}}

\newcommand{\be}{\begin{equation}}
\newcommand{\ee}{\end{equation}}
\newcommand{\bea}{\begin{eqnarray}}
\newcommand{\eea}{\end{eqnarray}}
\newcommand{\bean}{\begin{eqnarray*}}
\newcommand{\eean}{\end{eqnarray*}}

\newenvironment{acknowledgment}{{\flushleft \bf Acknowledgments:}}{}

\begin{document}

\title{Construction of a general semilocal exchange-correlation hole model: Application to 
nonempirical meta-GGA functionals}
\author{Lucian A. Constantin}
\affiliation{Center for Biomolecular Nanotechnologies (CBN) of the Italian Institute of Technology (IIT),
Via Barsanti, I-73010 Arnesano, Italy}
\author{Eduardo Fabiano}
\affiliation{National Nanotechnology Laboratory (NNL), Istituto di Nanoscienze-CNR,
Via per Arnesano 16, I-73100 Lecce, Italy }
\author{Fabio Della Sala}
\affiliation{National Nanotechnology Laboratory (NNL), Istituto di Nanoscienze-CNR,
Via per Arnesano 16, I-73100 Lecce, Italy }
\affiliation{Center for Biomolecular Nanotechnologies (CBN) of the Italian Institute of Technology (IIT),
Via Barsanti, I-73010 Arnesano, Italy}

\date{\today}

\begin{abstract}
Using a reverse-engineering method we 
construct a meta-generalized gradient approximation (meta-GGA) 
angle-averaged exchange-correlation hole model which has a general applicability. 
It satisfies known exact hole constraints and 
can exactly recover the exchange-correlation energy density of any reasonable 
meta-GGA exchange-correlation energy functional satisfying a minimal set of exact properties.
The hole model is applied to several non-empirical meta-GGA functionals:
the Tao-Perdew-Staroverov-Scuseria (TPSS), the revised TPSS (revTPSS) and the 
recently Balanced LOCalization (BLOC) meta-GGA of 
L.A. Constantin, E. Fabiano, and F. Della Sala,   
(J. Chem. Theory Comput. $\mathbf{9}$, 2256 (2013)).
The empirical M06-L meta-GGA functional is also considered.
Real-space analyses of atoms and ions as well as wave-vector analyses
of jellium surface energies, show that the meta-GGA hole models,
in particular the BLOC one, are very realistic and
can reproduce many features of benchmark  XC holes.
In addition, the BLOC hole model can be used to estimate with good accuracy
the Coulomb hole radius of small atoms and ions.
Thus, the proposed meta-GGA hole models provide a valuable tool to validate in detail 
existing meta-GGA functionals, and can be further used in the development of
DFT methods beyond the semilocal level of theory.
\end{abstract}

\maketitle

\section{Introduction}
\label{sec1}

Kohn-Sham (KS) ground-state density functional theory (DFT) \cite{KS,bookdft} 
can be considered the most used method in electronic calculations of quantum
chemistry and condensed matter physics. Its practical implementation
is based on approximations of the exchange-correlation (XC) energy ($E_{xc}$), 
which is a subject of intense research \cite{scus}.

The simplest functionals, beyond the local density approximation \cite{KS} (LDA), 
are  those based on the generalized gradient approximation (GGA), which are
constructed using the electron density and its gradient.
These functionals can achieve reasonable accuracy at a moderate 
computational cost \cite{mukappa,peverati12,tognetti10,bremond12}.
However, due to the extreme simplicity of the GGA level, 
no GGA functional can be simultaneously highly
accurate for atoms and solids \cite{ELCB}, even if recent GGAs 
can achieve broad applicability, being rather accurate for solids, 
molecules, surfaces, and interfaces \cite{peverati12,bremond12,pbeint,zeta1,zeta2}.

One step further can be achieved by considering
meta-generalized-gradient-approximations (meta-GGAs) \cite{scus,PKZB}
which use the positive KS kinetic energy density $\tau(\R)=1/2\sum_i|\nabla\psi_i(\R)|^2$
(with $\psi_i$ being the occupied KS orbitals) as an additional ingredient to the GGA ones.
Meta-GGAs are the most sophisticated semilocal functionals, incorporating important
exact conditions. They display an improved overall accuracy with respect to the GGA
functionals \cite{m06-l,BLOC,hao13}, still having almost the same attractive computational cost.
For these reasons, recently, numerous meta-GGA functionals have been proposed
\cite{m06-l,BLOC,TPSS,revTPSS,regTPSS,mggams,VT08,vsxc,m05,zhao08,M11L,bmk,br89,becke96,js}.

Nowadays, most semilocal XC functionals are constructed by providing a suitable model for the
exchange-correlation energy density. Non-empirical XC functionals base this model
on the satisfaction of several exact properties of the XC energy, in order to avoid
the use of empirical data and obtain a more robust physical construction.
However, the overall procedure may give rise to some fundamental questions,
because the XC energy density is not a well defined physical quantity \cite{burke98} and,
more importantly, its integral (i.e. the corresponding XC energy) is defined up to
a gauge \cite{gauge}. Therefore, the satisfaction of exact constrains by non-empirical 
functionals is not a sufficient condition to guarantee the ``exactness'' of the
underlying XC energy density model.
 
For these reasons it is often useful to consider the alternative definition
of the XC energy functional in terms of the XC hole \cite{parr_book,becke88}:
\begin{equation}\label{ee1}
E_{xc} = \sum_{\sigma\sigma'}\int\int 
n_{\sigma}(\R)\frac{\bar{n}_{xc}^{\sigma\sigma'}(\R,\R+\mathbf{u})}{2u}d\R d\mathbf{u}\ ,
\end{equation}
where $n_{\sigma}$ is the $\sigma$-spin density and $\bar{n}_{xc}^{\sigma\sigma'}(\R,\R+\mathbf{u})$ 
is the (coupling-constant averaged) XC hole ($\sigma'$)-spin-density at position $\R+\mathbf{u}$ 
around an electron of
spin $\sigma$ at position $\R$. Note that Eq. (\ref{ee1}) can be also decomposed,
as usual, into exchange and correlation contributions, by considering the 
partitioning of the XC hole into a Fermi and a correlation hole as
\begin{equation}
\bar{n}_{xc}^{\sigma\sigma'}(\R,\R+\mathbf{u}) = 
\bar{n}_x^{\sigma}(\R,\R+\mathbf{u})\delta_{\sigma\sigma'} + \bar{n}_c^{\sigma\sigma'}(\R,\R+\mathbf{u})\ .
\end{equation}

Separating the $\R$ and angular $\mathbf{u}$ integrations from the radial integration along $u$, Eq. 
(\ref{ee1}) provides the real-space, or $u$-space, analysis of the XC energy
\begin{equation}\label{e_real}
E_{xc} = 4\pi\sum_{\sigma}N_\sigma\int_{0}^{\infty}u^2\frac{\langle n_{xc}\rangle_{\sigma}(u)}{2u}du\ ,
\end{equation}
where $N_\sigma=\int n_\sigma(\R)d\R$ and the angle- and system-averaged XC hole has been defined
as
\begin{equation}
\langle n_{xc}\rangle_{\sigma}(u) = \frac{1}{N_\sigma}\int d\R 
n_{\sigma}(\R)\int\frac{d\Omega_u}{4\pi}\sum_{\sigma'}\bar{n}_{xc}^{\sigma\sigma'}(\R,\R+\mathbf{u})\ ,
\end{equation}
with $d\Omega_u$ the solid angle element in the $\mathbf{u}$-space.
Introducing the Fourier transform of $1/u$ the wave-vector,
or $k$-space, analysis is obtained as \cite{LP}
\begin{equation}\label{wva_eq}
E_{xc} = \frac{1}{\pi}\sum_\sigma N_\sigma\int_0^\infty\langle n_{xc}\rangle_\sigma(k)d k\ .
\end{equation}
This is a very powerful tool for understanding in detail 
the behavior of density functionals 
at jellium surfaces\cite{js,PCP,CPP}, where the small- and large-$k$ 
asymptote are accurately (exactly) known.

The modeling of the angle- and system-averaged XC hole is a very important 
topic in DFT because, unlike for the energy density, the angle- and 
system-averaged XC hole is uniquely defined and is an observable (at full coupling strength) \cite{PBW}.
Therefore, its knowledge can be used via Eq. (\ref{e_real}) to
construct non-empirical physically-motivated XC functionals \cite{br89,becke88,PBW}.
Alternatively, the modeling of the angle- and system-averaged XC hole through reverse
engineering from an already existing XC approximation can be employed to check the
internal consistency of the functional, by comparison with known
exact properties of the XC hole. In fact, any XC functional is better described 
by its underlying hole, which analyzes the XC energy into 
contributions of various distances $u$ from an electron.  

The construction of the XC hole model is well established at the LDA 
\cite{PW,EP,a112} and GGA \cite{CPP,PBW,EP,Scush,Scush1} levels. However, concerning 
meta-GGA functionals, to our knowledge, only the TPSS angle- and system-averaged
XC hole model has been back engineered \cite{CPT}, proving to be
among the most realistic ones currently available in
density functional theory.

In this paper we consider the problem of constructing realistic hole models for
sophisticated XC functionals in closer details. Thus, we provide a general
model for the angle- and system-averaged XC hole. The proposed construction
can be used for any XC functional satisfying a minimal set of exact
constraints of the XC energy. Our model is then applied to construct the
angle- and system-averaged XC holes of the non-empirical TPSS \cite{TPSS}, 
revTPSS \cite{revTPSS}, and BLOC \cite{BLOC} meta-GGA functionals,
which are accurate semilocal functionals respecting many exact constraints of the
XC energy, as well as of the widely used empirical M06-L exchange-correlation functional \cite{m06-l}.
Finally, the TPSS, revTPSS, and BLOC hole models are applied to perform a real-space 
analysis of the XC energy of several small atoms and ions, to probe the 
reliability of the Fermi and Coulomb holes in each case.

\section{Construction of the XC hole model}
\label{secii}
In this section we outline the construction of a general XC hole model
for any functional respecting minimal constraints of the exchange and correlation
energy.

\subsection{Exchange hole}
The procedure for the exchange part closely follows that of Ref. \onlinecite{CPT}
and is briefly repeated here for sake of completeness and to underline better the generality of this
construction.
The spin-scaling relation of the exchange energy hole \cite{PBW}:
\begin{equation}
\bar{n}_{x}[n_{\uparrow} ,n_{\downarrow}](\mathbf{r},\mathbf{r}+\mathbf{u})=
\sum_{\sigma}\frac{n_{\sigma}(\mathbf{r})}{n(\mathbf{r})}\bar{n}_{x}[2
n_{\sigma}](\mathbf{r},\mathbf{r}+\mathbf{u}),
\label{e7b}
\end{equation}
allows to consider only the spin-unpolarized case.

We consider an exchange functional of the form
\begin{equation}
E_x^{MGGA} = \int n \epsilon_x^{LDA}(\R)F_x^{MGGA}(s(\R),z(\R))d\R\ ,
\end{equation}
where $\epsilon_x^{LDA}=-(3/4)(3/\pi)^{1/3}n^{1/3}$ is the LDA exchange energy per particle,
$s=|\nabla n|/2nk_F$ is the reduced gradient for exchange ($k_F=(3\pi^2n)^{1/3}$),
and $z=\tau^W/\tau$ ($0\leq z\leq 1$) is
the meta-GGA ingredient \cite{TPSS} that can indicate the iso-orbital
regions (when $z\rightarrow 1$), and the slowly-varying regime (when $z \rightarrow 0$),
with $\tau^W=|\nabla n|^2/(8n)$ being the von Weizs\"{a}cker kinetic energy density
\cite{bookdft}.
We assume that the functional fulfills the following constraints:
\begin{itemize}
\item[(i)] It recovers LDA for a homogeneous electron gas (i.e. $F_x^{MGGA}=1$ for $s=0$ and $z=0$);
\item[(ii)] It can describe accurately one-electron densities.
\end{itemize}
In addition we ask the functional to be rather accurate for any problem ranging from atoms to
jellium models.
Then, following Ref. \onlinecite{CPT}, we use the following ansatz for the
angle- and system-averaged exchange hole (for simplicity the spin index was dropped):
\begin{equation}
\langle n_x^{MGGA}\rangle (u) = \frac{1}{N}\int n^2(\R)J^{MGGA}[s(\R),z(\R),y(\R)]d\R\ ,
\end{equation}
where $y=k_F(\R)u$, and
\begin{eqnarray}
\nonumber
J^{MGGA} & = & \Bigg\{-\frac{\mathcal{A}}{y^2}\frac{1}{1+(4/9)\mathcal{A}y^2} + 
\Bigg[\frac{\mathcal{A}}{y^2} + \mathcal{B} +  \\
\nonumber
&&  +\mathcal{C}\left(1+s^2\mathcal{F}(s,z)\right)y^2 + 
\mathcal{E}\left(1+s^2\mathcal{G}(s,z)\right)y^4 +  \\
\label{ej}
&&  +s^2\mathcal{K}(s,z)y^6\Bigg]e^{-\mathcal{D}y^2}\Bigg\}e^{-s^2\mathcal{H}(s,z)y^2}\ .
\end{eqnarray}

The parameters $\mathcal{A}$, $\mathcal{B}$, $\mathcal{C}$, $\mathcal{D}$, and $\mathcal{E}$ can
be fixed considering that condition (i) implies that for $s=0$ the hole model must recover the
LDA exchange hole. Thus, comparing with Eq. (18) of Ref. \onlinecite{EP}, we can fix
$\mathcal{A}=1.0161144$, $\mathcal{B}=-0.37170836$, $\mathcal{C}=-0.077215461$, 
$\mathcal{D}=0.57786348$, and $\mathcal{E}=-0.051955731$.
Note that this choice also ensures that the present exchange hole model satisfies
the exact constraint for the on-top exchange hole \cite{EP}, that is
$J^{MGGA}(s,z,y=0)=J^{LDA}(0)=-1/2$.

The three functions $\mathcal{F}$, $\mathcal{G}$, and $\mathcal{K}$ can be fixed instead by 
requiring the satisfaction of the following conditions:
\begin{itemize}
\item[(a)] \textbf{Energy sum rule} of Eq (\ref{e_real}), which leads to the integral equation
\begin{equation}
\frac{8}{9}\int_0^\infty J^{MGGA}(s,z,y)ydy = -F_x^{MGGA}(s,z)\ ;
\end{equation}
\item[(aa)] \textbf{Particle sum rule}, $\int n_x(\mathbf{u},\R+\mathbf{u})d\mathbf{u}=-1$,
which is the normalization condition of the exchange hole. This constraint gives the integral equation
\begin{equation}
\frac{4}{3\pi}\int_0^\infty J^{MGGA}(s,z,y)y^2dy = -1\ ;
\end{equation}
\item[(aaa)] \textbf{Small-$u$ behavior}. Any accurate DFT functional must describe well
the local  (nearsighted) effects \cite{prodan05}. 
Thus, the hole model must have the correct small-$u$ expansion
as in Eq. (26) of Ref. \onlinecite{CPT}. This implies the differential condition
\begin{equation}
\frac{\partial^2 J^{MGGA}(s,z,y)}{\partial y^2}\Big|_{y=0} = 
\frac{1}{3}\frac{s^2}{u(s)}\left(\frac{1}{z}+1\right)\ ,
\end{equation}
with $u(s)=1$ or $u(s)=(1+cs^6)^{1/3}$ with $c=0.00012$, as discussed in Ref. \onlinecite{CPT}.
\end{itemize}
Using Eq. (\ref{ej}) the conditions (a)-(aaa) provide a set of three linear equations that completely
define the functions $\mathcal{F}$, $\mathcal{G}$, and $\mathcal{K}$. These are different for
each specific functional (they depend on $F_x^{MGGA}$).

Finally, the function $\mathcal{H}$ can be approximated by the flexible ansatz
\begin{equation}
\mathcal{H}(s,z) = z^\alpha\frac{\sum_i^{N_1}a_is^{i-1}}{1+\sum_j^{N_1+1}b_is^{j}}\ ,
\end{equation}
where parameters $a_i$ and $b_j$ are fitted, in agreement with condition (ii), to reproduce 
the exchange hole of reference one-electron densities (e.g., hydrogen and Gaussian densities,
which are relevant for quantum chemistry \cite{zeta1,zeta2,scal}). The parameter $\alpha$
can instead be obtained by fitting the wave-vector analysis of jellium surfaces.

The present exchange hole is of general utility and can be applied to any
reasonably accurate functional, including all the TPPS-like exchange functionals
\cite{TPSS,revTPSS,BLOC}, VT\{8,4\} \cite{VT08}, and the new meta-GGA of Sun et al.
\cite{mggams}.

\subsection{Correlation hole}
\label{corh_sec}
For the correlation part we focus on a spin- and angle-averaged hole model,
defined as
\begin{equation}
\bar{n}_c^{MGGA}(\R,u) = \sum_{\sigma\sigma'} \frac{n_\sigma(\R)}{n(\R)}\int
\bar{n}_{c}^{\sigma\sigma'}(\R,\R+\mathbf{u})\frac{d\Omega_u}{4\pi} \ .
\end{equation}
The use of a spin-averaged model is in fact more general because it can be used both for functionals
where the spin resolution is known and for those (like TPSS, revTPPS, and BLOC) where instead it is 
not defined.

We consider any semilocal correlation functional of the form
\begin{equation}
E_c^{MGGA} = \int 
n(\R)\epsilon_c^{MGGA}(\UP,\DN,\nabla\UP,\nabla\DN,\TUP,\TDN)d\R\ ,
\end{equation}
and we assume that the functional is reasonably
accurate for any problem in electronic theory and 
\begin{itemize}
\item [(b)] it recovers LDA for the homogeneous electron gas (i.e. when $\UP$ and $\DN$ are constant 
over the space);
\item [(bb)] it is one-electron self-interaction free.
\end{itemize}
Such a functional can be described by the following general reverse-engineering
spin- and angle-averaged correlation hole model
\begin{eqnarray}
\nonumber
\bar{n}_c^{MGGA}(\R,u) & = & \bar{n}_c^{MGGA}[r_s(\R),\zeta(\R),t(\R),z(\R)](v) = \\
\nonumber
& = & \phi^5k_s^2\Big[A_c(r_s,\zeta,v)\mathrm{sgn}(1-z|\zeta|) + \\
\label{enc}
&& + t^2B_c^{MGGA}(r_s,\zeta,t,z,v)\Big]\theta(v_c-v)\ ,
\end{eqnarray}
where $r_s=[3/(4\pi n)]^{1/3}$ is the local Seitz radius,
$\zeta=(n_\uparrow-n_\downarrow)/n$ is the relative spin-polarization,
$\phi=((1+\zeta)^{2/3}+(1-\zeta)^{2/3})/2$ is a spin-scaling factor, 
$t=|\nabla n|/2n k_s\phi$
($k_s=(4k_F/\pi)^{1/2}$ being the Thomas-Fermi screening wave vector), and
$v=\phi k_s u$ is the reduced electron-electron separation on the scale
of the screening length.
The function $A_c$ is the one of Refs. \onlinecite{PBW,PW}, such that
$\phi^5k_s^2A_c$ is the LDA correlation hole. Hence, condition (b) is fulfilled. 
The function $B_c$ is chosen in analogy with Ref. \onlinecite{PBW} to be
\begin{eqnarray}
\nonumber
B_c^{MGGA}(r_s,\zeta,t,z,v) & = & 
B_c^{LM}(v)\mathrm{sgn}(1-z|\zeta|)\left[1-e^{-\eta^{p_1}}\right] + \\
&& + \mu^{MGGA}(r_s,\zeta,t,z)v^2e^{-\eta^{p_2}}\ .
\end{eqnarray}
Here $B_c^{LM}(v)$ is the RPA nonoscilating long-range contribution
(see Eq. (49) of Ref. \onlinecite{PBW}), $\eta=\sqrt{p}v$ is
the scaled distance suitable for the gradient correction \cite{PBW},
with $p(r_s,\zeta)=\pi k_F(0.305-0.136\zeta^2)/4\phi^4$ measuring where
the short range contribution vanishes. The parameters $p_1\geq 2$ and $1<p_2\leq2$
control the balancing between the long- and short-range
behavior. They are set both to 2 in Ref. \onlinecite{PBW}. We found instead
that a better balancing can be found by using $p_1=3$ and $p_2=3/2$.
We remark however that the shape of the correlation hole is weakly dependent
on the choice of these two parameters.

The function $\mu^{MGGA}$ is fixed by imposing the energy sum rule 
$2\pi \int \bar{n}_c^{MGGA}(\R,u)u du=\epsilon_c^{MGGA}(\R)$. It is
\begin{eqnarray}
\nonumber
&&\mu^{MGGA}(r_s,\zeta,t,z) = \Bigg[\frac{\phi^2k_s^2}{2\pi}\epsilon_c^{MGGA} + \\
\nonumber
&&\quad - k_s^2\phi^5\mathrm{sgn}(1-z|\zeta|)\int_0^{v_c}A_cvdv + \\
\nonumber
&& \quad -k_s^2\phi^5t^2\mathrm{sgn}(1-z|\zeta|)\int_0^{v_c}B_c^{LM}(1-e^{-\eta^{p_1}})vdv\Bigg]\Bigg/ \\
&& \quad \Bigg/ \Big[k_s^2\phi^5t^2\int_0^{v_c}e^{-\eta^{p_2}}v^3dv\Big]\ .
\end{eqnarray}
This function controls the short-range (small $v$) behavior
of the hole model, which is the most important for a semilocal hole. 
In contrast with the case of Ref. \onlinecite{PBW}
it is not constructed from the slowly-varying behavior of any underlying functional,
but it is instead entirely determined by the energy sum rule. Therefore,
it is more general and can apply to any correlation functional.

Finally, as for Ref. \onlinecite{PBW}, the cutoff parameter $v_c$ is chosen as the largest root
that enforces the particle sum rule $\int \bar{n}_c^{MGGA}(\R,u)d\mathbf{u}=0$.
Note that $\mu^{MGGA}$ is dependent on $v_c$.

The correlation hole model of Eq. (\ref{enc}) has the following properties:
\begin{enumerate}
\item By construction it recovers the LDA correlation hole for a homogeneous electron gas 
(condition (b));
\item $\bar{n}_c^{MGGA}(\R,u=0)$ recovers the accurate LDA on-top hole \cite{BEP};
\item It recovers the LDA cusp, i.e. $d\bar{n}_c^{MGGA}/dv|_{v=0} = d\bar{n}_c^{LDA}/dv|_{v=0}$.
Note that the value of $p_2$ is not altering the description of the cusp but just modulating
the short-range hole behavior;
\item By construction it satisfies the particle and energy sum rules (note that instead the PBE and TPSS
hole models of Ref. \onlinecite{PBW} and \onlinecite{CPT} do not recover exactly the PBE and TPSS 
correlation 
energy densities);
\item It has the correct one-electron and iso-orbital limit (condition (bb)). This is ensured by the 
function $\mathrm{sgn}(1-z|\zeta|)$, which is always 1 except in the iso-orbital regime, when $z=1$ 
and $\zeta=1$, in which case it is 0.
\item In the slowly-varying ($t\leq 1$) long-range ($0\ll v\leq v_c$) limit, the hole model
recovers the semilocal RPA nonoscilatory behavior, which is the right limit for any reasonable
GGA and meta-GGA functional \cite{PBW,PCP}.
\end{enumerate}

The present correlation hole model has a simple and numerically robust expression that can be easily 
implemented in any code. It uses the cutoff procedure only once, and thus it has almost the same 
computational cost as the GGA hole.  
On the other hand, the TPSS correlation hole of Ref. \onlinecite{CPT}
is not expressed in a general form and it is only suitable for correlation functionals
having the TPSS correlation formula. Furthermore, it contains terms as 
$\frac{\epsilon^{GGA}_{c}(n_{\uparrow},n_{\downarrow},\nabla n_{\uparrow},\nabla
n_{\downarrow})}{\epsilon^{GGA}_{c}(n,0,\nabla n,0)}$, where
$\epsilon^{GGA}_{c}$ is the correlation energy density of the GGA functional underlying the
meta-GGA correlation (e.g. for TPSS, $GGA=PBE$). Such an expression may
lead to numerical instabilities when the GGA correlation used to construct the meta-GGA
functional is rapidly vanishing in the tail (as in the BLOC case \cite{BLOC}).

\section{Construction of the TPSS, revTPSS, and BLOC XC holes}
In this section we apply the XC hole model developed above to outline
the construction of the XC holes of the nonempirical TPSS \cite{TPSS}, revTPSS
\cite{revTPSS}, and BLOC \cite{BLOC} meta-GGA functionals. These functionals
satisfy numerous exact constraints of the XC energy and are thus suitable for
an investigation of their corresponding XC holes. 

Concerning the exchange part, all the three functionals share the same general form 
of the enhancement factor \cite{TPSS,revTPSS,BLOC}
\begin{equation}
F_x=1+\kappa -\kappa/(1+x/\kappa)\ ,
\label{e2}
\end{equation}
with $\kappa=0.804$ and
\begin{eqnarray}
x=\{ [\frac{10}{81}+c\frac{z^f}{(1+z^2)^2}]s^2+\frac{146}{2025}\tilde{q}_b^2  \nonumber 
\\
-\frac{73}{405}\tilde{q}_b \sqrt{\frac{1}{2}(\frac{3}{5}z)^2+\frac{1}{2}s^4}+
\frac{1}{\kappa}(\frac{10}{81})^2 s^4 \nonumber \\
+2\sqrt{e}\frac{10}{81}(\frac{3}{5}z)^2 +
e\mu s^6    \} / (1+\sqrt{e}s^2)^2,
\label{e3}
\end{eqnarray}
where $\tilde{q}_b=(9/20)(\alpha-1)/\sqrt{1+0.4 \alpha(\alpha-1)}+2s^2/3$ 
mimics the reduced Laplacian of the density
$q=\nabla n^{2}/\{4(3\pi^{2})^{2/3}n^{5/3}\}$, with $\alpha=(5/3)s^2(1/z-1)$;
and the coefficients $c$, $f$, $e$, and $\mu$ are shown in the Table \ref{tab1}.
%
\begin{table}
 \begin{center}
 \caption{\label{tab1}Coefficients of the TPSS, revTPSS and BLOC
 exchange meta-GGA functionals.}
 \begin{ruledtabular}
 \begin{tabular}{lrrrr}
 Functional & \multicolumn{1}{c}{c} & \multicolumn{1}{c}{f} &
 \multicolumn{1}{c}{e} & \multicolumn{1}{c}{$\mu$} \\
 \hline
 TPSS      & 1.59096 & 2         & 1.5370 &  0.21951 \\
 revTPSS   & 2.35204 & 3         & 2.1677 &  0.14000 \\
 BLOC      & 1.59096 & $4-3.3 z$ & 1.5370 &  0.21951 \\
 \end{tabular}
 \end{ruledtabular}
 \end{center}
 \end{table}
%
For this reason, their exchange hole basically coincides with the
TPSS one of Ref. \onlinecite{CPT}, except for the use,
for each functional, of different parameters in the enhancement factor
employed to determine the functions $\mathcal{F}$, $\mathcal{G}$, and $\mathcal{K}$.

\begin{figure}
\includegraphics[width=\columnwidth]{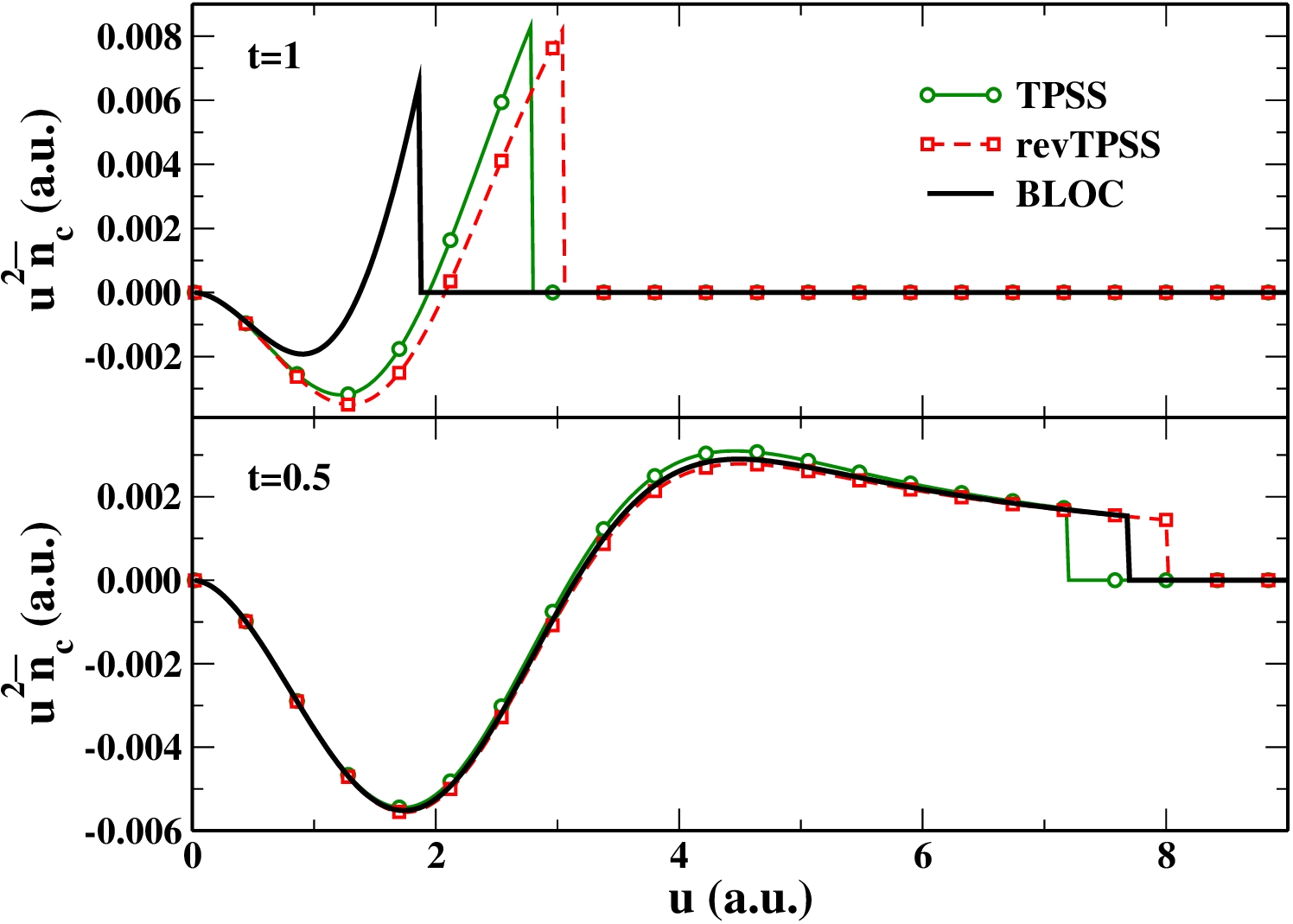}
\caption{\label{fig_chole} Comparison of the TPSS, revTPSS, and BLOC angle-averaged correlation holes 
for $r_s=2$, $\zeta=0$, $z=0.5$, and $t=0.5$ and 1, respectively.} 
\end{figure}
To construct the correlation hole according the reverse engineering model
of Section \ref{corh_sec}, only the correlation energy per particle is required.
This has a TPSS-like form \cite{TPSS} for all the three functionals considered here, mainly
differing only for the choice of the correlation parameter: 
it is fixed to $\beta=0.066725$ for TPSS; 
it is a fit of the Hu and Langreth expression
\cite{hu86} $\beta=\beta(r_s)$ for revTPSS; and for BLOC, $\beta$ is defined as \cite{TPSSloc}
\begin{equation}
\beta(r_s,t)=\beta_0+a\;t^2 (1-e^{-r_s^2})\ ,
\end{equation} 
with $\beta_0=0.0375$ obtained from the LDA linear response, 
and $a=0.08$ obtained by
jellium surface analysis. This choice provides a stronger
localization of the correlation energy density, cutting spurious
contribution in the valence and tail region \cite{TPSSloc}.
We remark once more that, because of this stronger localization, the 
construction of the BLOC correlation hole would be problematic with the
model proposed in Ref. \onlinecite{CPT}, because the correlation energy
per particle vanishes very rapidly in the tail of the density.

A comparison of the TPSS, revTPSS, and BLOC angle-averaged correlation
holes is reported in Fig. \ref{fig_chole}, for $r_s=2$, $\zeta=0$, $z=0.5$ and
$t=0.5$ and 1, respectively.
For all functionals, the plot of the hole is rather similar,
reflecting the similarities  in the functionals' construction.
At larger values of $t$, however, the BLOC hole starts to
show a clear tendency towards a higher degree of
localization, in agreement with the paradigm underlying 
the development of the BLOC (TPSSloc) correlation functional \cite{TPSSloc}.

\begin{figure}
\includegraphics[width=\columnwidth]{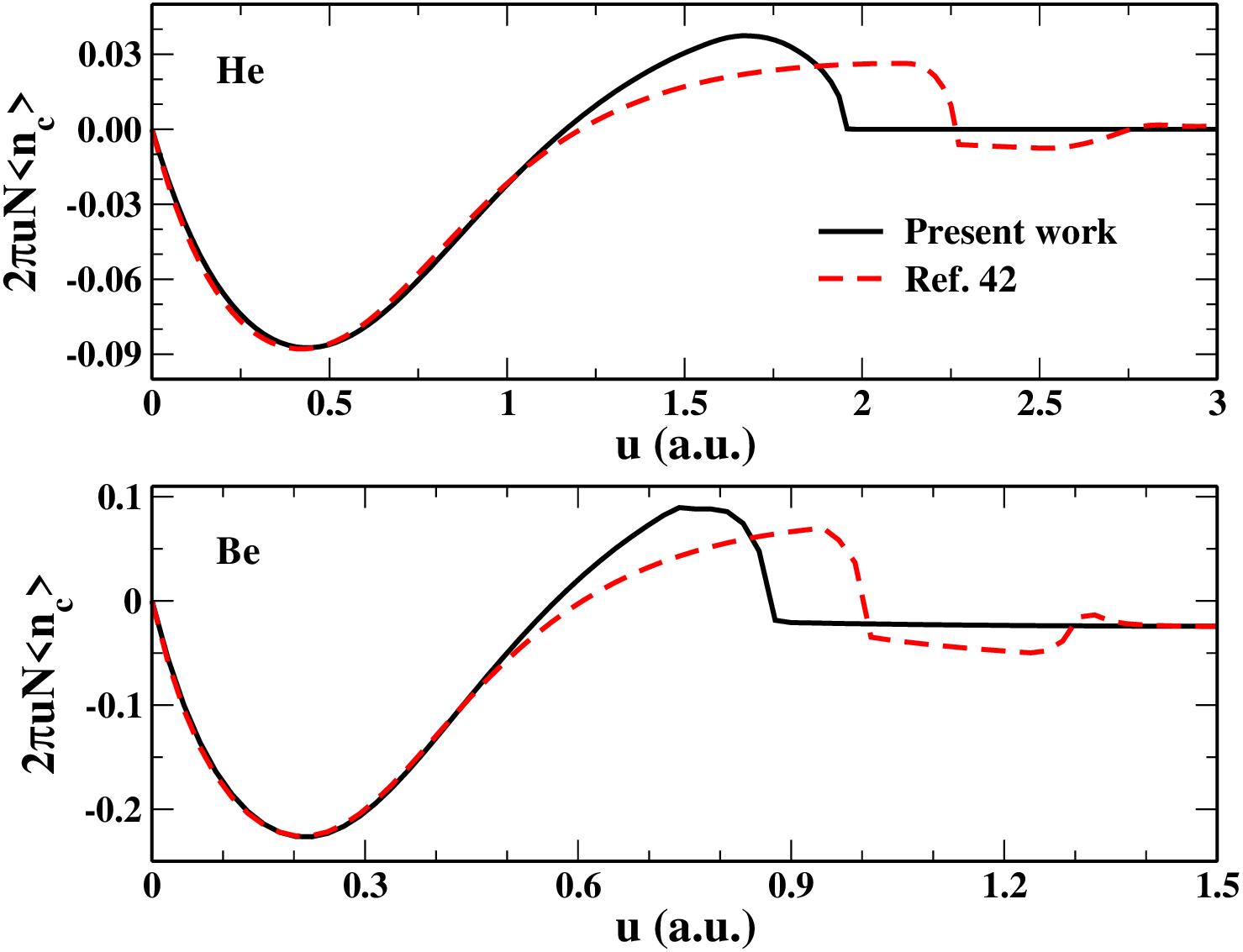}
\caption{\label{comp_fig} Comparison of the real space analyses of the TPSS correlation holes 
obtained with the present construction and with the method of Ref. \onlinecite{CPT}, for the He and 
Be atoms.} 
\end{figure}
Finally, in Fig. \ref{comp_fig} we compare the real space analyses of the TPSS correlation holes obtained 
with the present construction and with the method of Ref. \onlinecite{CPT}, for the He and Be atoms.
Both constructions agree very well for small and medium values of $u$ (i.e. inside the Coulomb
hole radius; see Tab. \ref{tab_rc}), that is the region contributing to most of the correlation energy.
For large $u$-values the hole model introduced in this work goes to the nonoscillatory semilocal
RPA hole, whereas the model of Ref. \onlinecite{CPT} shows a more structured behavior due to the repeated 
use of the cutoff procedure.

\section{Wave vector analysis of jellium surface energies}
As a first application of the constructed hole models,
we study the wave vector analysis of jellium surface energies (see Eq. (\ref{wva_eq})).
All calculations were done similarly as in Refs. \onlinecite{js,PCP,CPP,CPT},
using accurate LDA orbitals.
\begin{figure}
\includegraphics[width=\columnwidth]{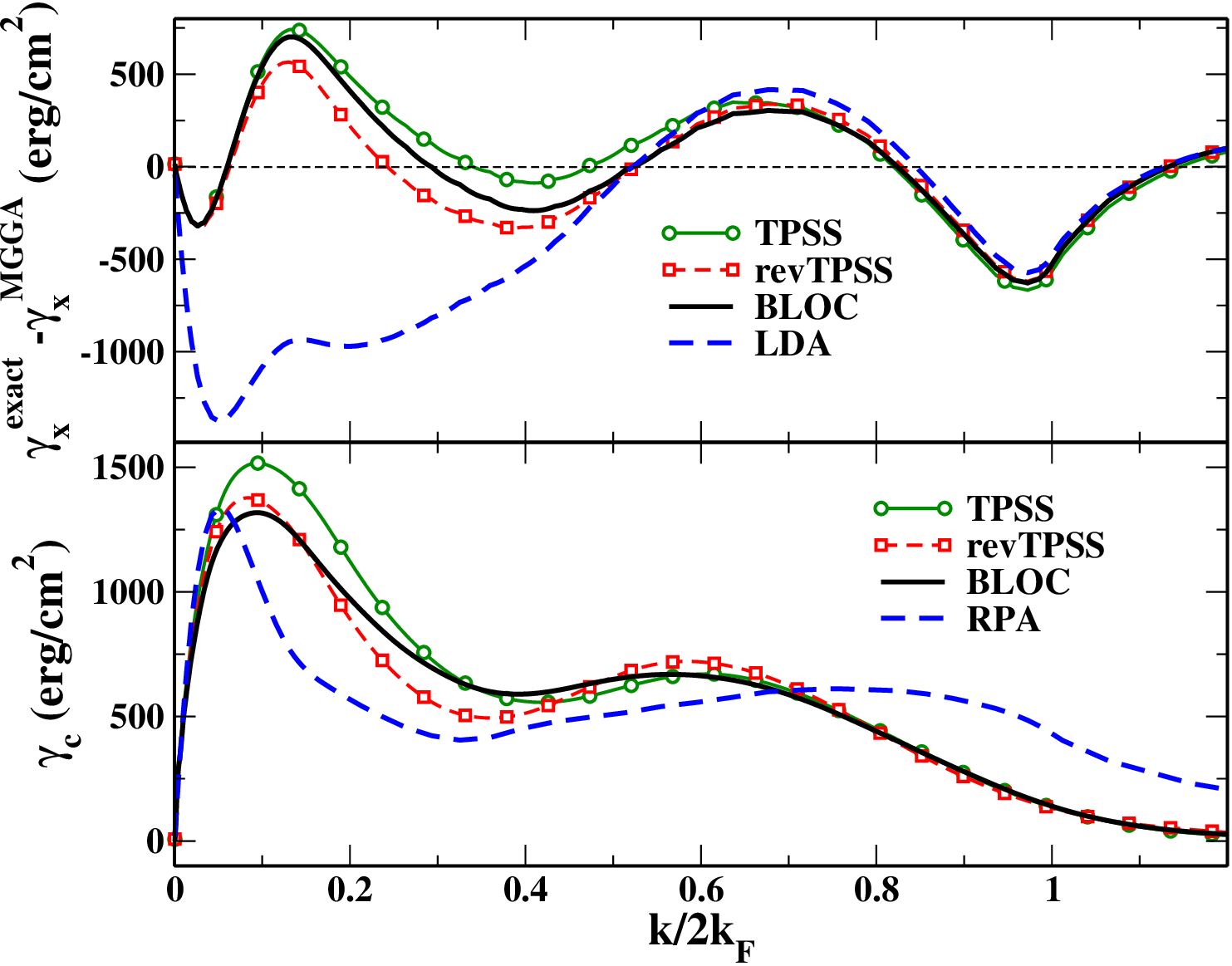}
\caption{\label{wave_fig}Wave vector analysis of jellium surface energy for a jellium slab of 
thickness $a=2.23\lambda_F$ and $r_s=2.07$ as resulting from the TPSS, revTPSS, and BLOC hole models: 
upper panel, errors on exchange contribution (LDA is also reported for reference); lower panel, 
correlation contribution (RPA is also reported as reference).} 
\end{figure}

In the upper panel of Fig. \ref{wave_fig}, we report the errors in the wave vector analysis 
of exchange jellium surface energies. 
For $k\geq k_F$, all the meta-GGAs recover the LDA behavior, becoming eventually
exact for large values of $k$ \cite{LP}.
At smaller values of $k$ instead LDA becomes highly inaccurate, and the meta-GGA holes show
a strong improvement for this plasmonic region. In particular, at quite small values
of $k$, revTPSS is the best functional, whereas at medium values of $k$, TPSS yields
the smallest error. The BLOC results, being always in between the other two, provide 
thus the best compromise for $k\leq k_F$. These results are in agreement with the findings
of Refs. \onlinecite{revTPSS,BLOC} where BLOC and revTPSS were found to be well accurate for jellium
exchange energies.

The results for correlation surface energies are reported in the lower panel of Fig. \ref{wave_fig}.
The plot shows that in the plasmonic region ($k\rightarrow 0$) all the hole models 
are remarkably accurate, reproducing the correct RPA behavior, thanks to 
the non-oscillatory long-range RPA gradient correction term ($B_c^{LM}$) present
in the hole expressions. At the same time, for large-$k$ values, they recover
the correct LDA behavior. Note that instead RPA fails in this region.
We recall that revTPSS and BLOC XC and correlation-only jellium surface energies are
very accurate \cite{TPSSloc,BLOC,revTPSS}, and thus the error cancellation between exchange
and correlation is minimal, in contrast to TPSS that relays on an error cancellation between exchange 
and correlation.

\section{Real space analysis of atoms}
The exchange and correlation hole models of the TPSS, revTPSS, and BLOC functionals
are applied to the real space analyses of several atoms and ions, for which accurate
benchmark results are available, in order to assess the 
physical content of the
functionals with respect to their ability to accurately describe the XC hole.
We recall in fact that the XC hole is an observable at full coupling strength \cite{PBW} and,
differently from the energy density, it is uniquely defined for each system and functional.
We also stress that the real-space analysis of the individual 
exchange and correlation contributions
provides a much deeper inside into the physical content of different functionals than the
conventional assessment based on the comparison of integrated XC (or total) energies.
In fact, the latter incorporates important error cancellation effects. Thus, 
it may happen that functionals displaying poor exchange hole behavior, provide 
quite accurate integrated XC results.

The calculation of the atomic hole functions have been performed using analytic Hartree-Fock
orbitals and densities \cite{CR11}. For the Hookium instead exact orbitals were
employed \cite{hookium}. We recall that the Hookium \cite{js,TPSSloc,hookium,hooke1,hooke2,hooke3}
represents two interacting electrons in an isotropic harmonic potential with the
harmonic force constant $k=1/4$ \cite{hookium}.

\subsection{Exchange hole}
\begin{figure}
\includegraphics[width=\columnwidth]{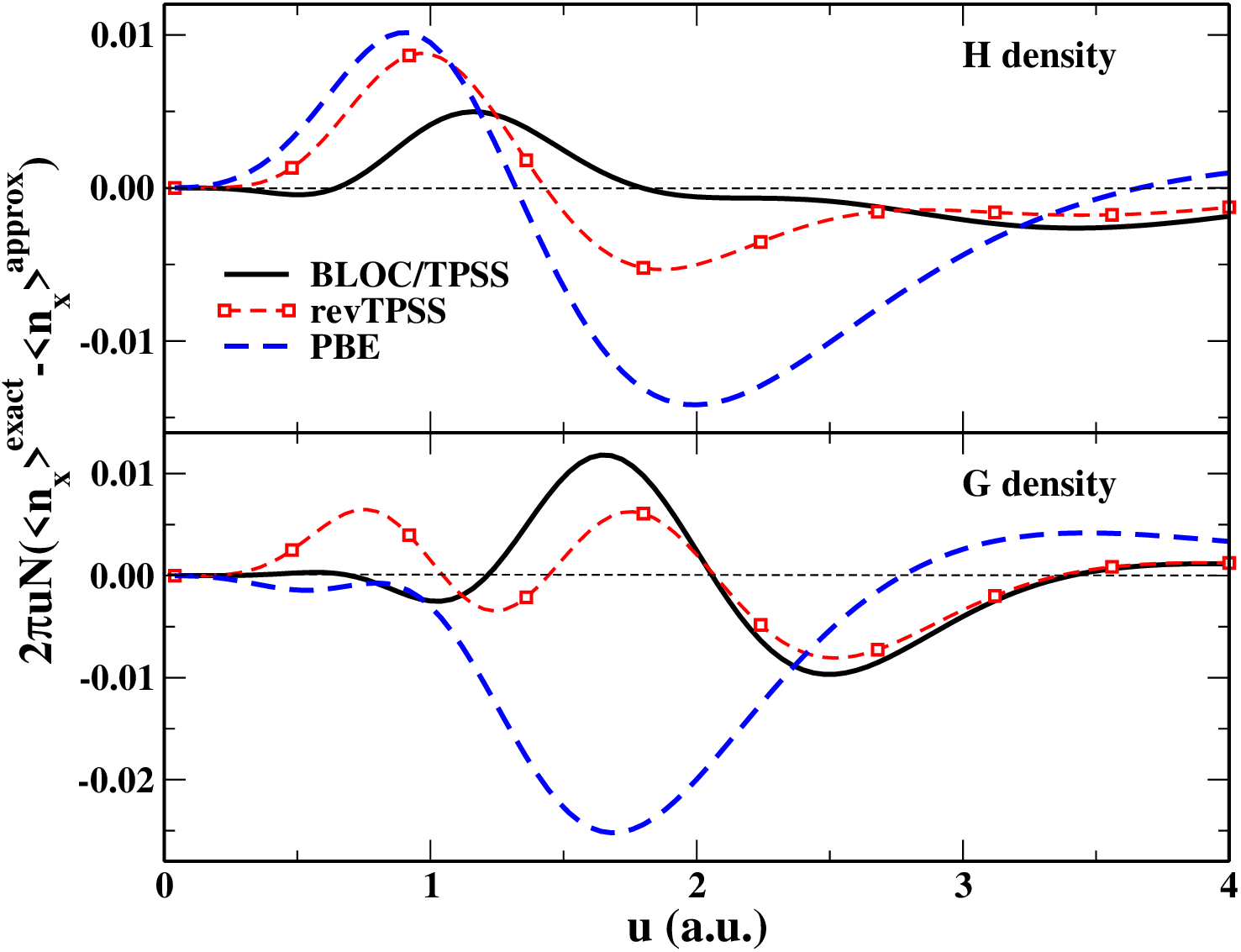}
\caption{\label{hg_xhole_fig}Errors in the real space analysis of the exchange energy of the hydrogen 
(upper panel) and Gaussian (lower panel) one-electron densities.
Exact values are reported in Ref. \onlinecite{CPT}.} 
\end{figure}
In Fig. \ref{hg_xhole_fig}, we report the errors in the real-space analysis ($2\pi N u 
(\langle\bar{n}^{exact}_{x}\rangle(u)-\langle\bar{n}^{approx}_{x}\rangle(u))$) 
for the hydrogen and Gaussian one-electron densities. 
These systems are important because they are used as model systems in the development of the meta-GGA
functionals \cite{PTSS} and hole models \cite{CPT} (through the function 
$\mathcal{H}(s,z)$). Their exact holes 
are analytical \cite{CPT}. Therefore, their real-space 
analysis provides an 
interesting validation for 
the exchange hole model. We recall that for one-electron densities the BLOC meta-GGA coincides with 
TPSS, 
by construction. Thus, only BLOC and revTPSS are reported. The PBE hole model of Ref. \onlinecite{EP} 
is also shown for comparison.

Inspection of the plot shows that, indeed, revTPSS and BLOC/TPSS improve considerably over PBE, 
for both model densities. In particular, in the case of the hydrogen density, BLOC/TPSS is the most 
accurate for $u\leq 2.5$, while  revTPSS is slightly better at large distances ($u\geq 2.5$). 
This is in accord with previous work \cite{revTPSS}, where it has been shown that the TPSS exchange 
energy density is closer to the exact conventional one, than the revTPSS one.
For the Gaussian density, again the TPSS hole is better at smaller $u$, while revTPSS yields the 
smallest error at larger $u$. Nevertheless, both TPSS and revTPSS curves are accurate, with similar 
performance.

\begin{figure}
\includegraphics[width=\columnwidth]{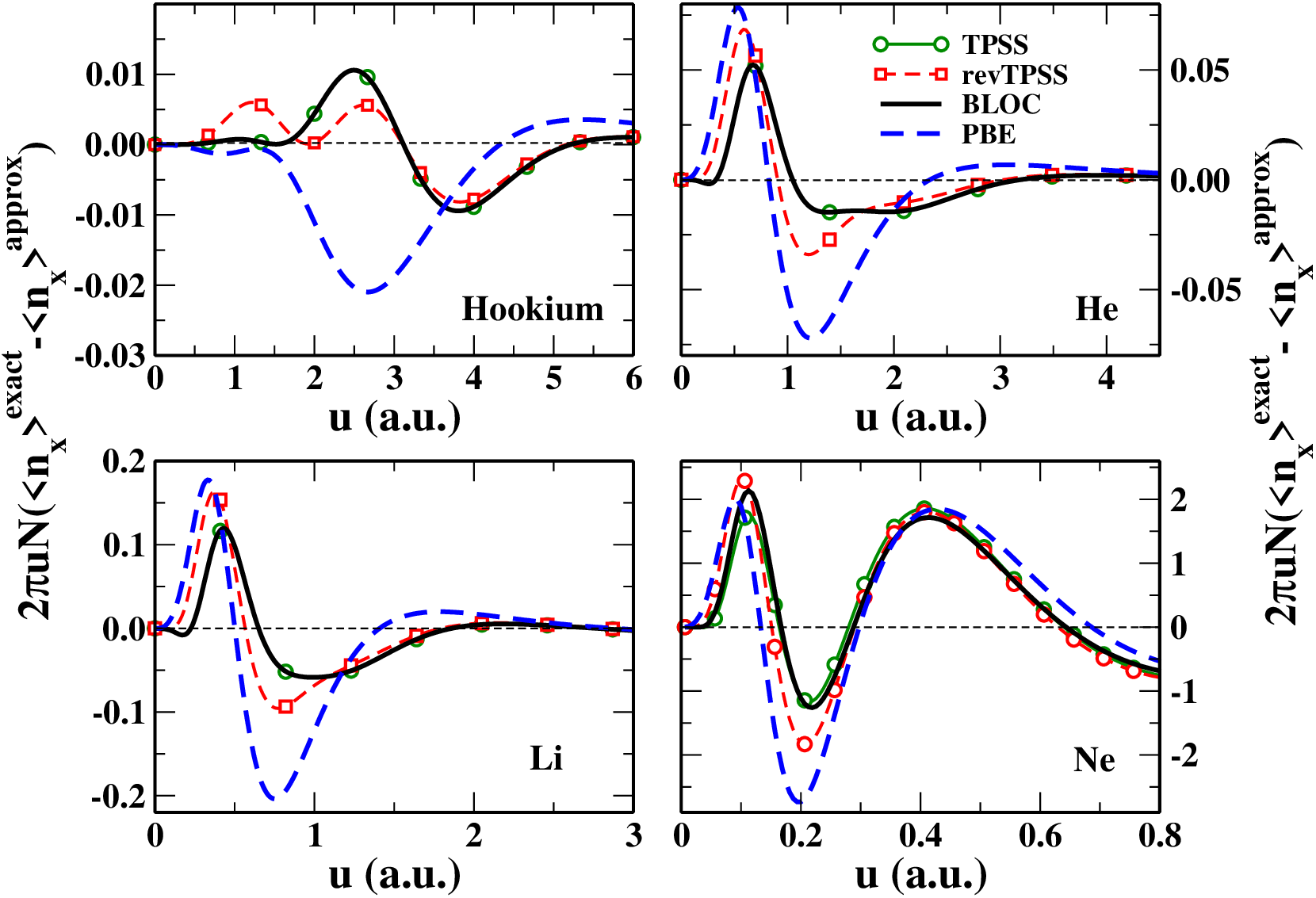}
\caption{\label{xhole_atoms_fig}Errors in the real space analysis of the exchange energy of several 
atoms. Note that for two-electron atoms TPSS and BLOC coincide by construction.} 
\end{figure}
The errors on the real-space analysis of the exchange energy of several atoms were also considered and
are shown in Fig. \ref{xhole_atoms_fig}. We remark that for two-electron systems the BLOC exchange 
coincides with the TPSS one by construction, while in the Li and Ne atoms, where the density varies
quite rapidly (i.e.  $z\geq 0.6$), in most regions BLOC and TPSS are very close to each other.
The plotted curves confirm the observations made for the model densities and indicate that all the meta-GGA 
functionals provide sensibly improved exchange holes with respect to PBE. These results
however suggest that the improvement of meta-GGA exchange holes is especially important 
for more diffuse densities (e.g. the Hookium and light atoms), whereas in the Ne atom the differences
are less marked. Anyway, a very close behavior is found in large-$u$ regions for all the functionals, 
because all the exchange holes considered recover the same non-oscillatory exponential decay (dictated 
by the function $\mathcal{H}$) in that limit.
Thus, the large-$u$ behavior may supposed to be less accurate (it is however also less important
for the energy).

\subsection{Correlation hole}
\label{corr_hole_sec}

\begin{figure}
\includegraphics[width=\columnwidth]{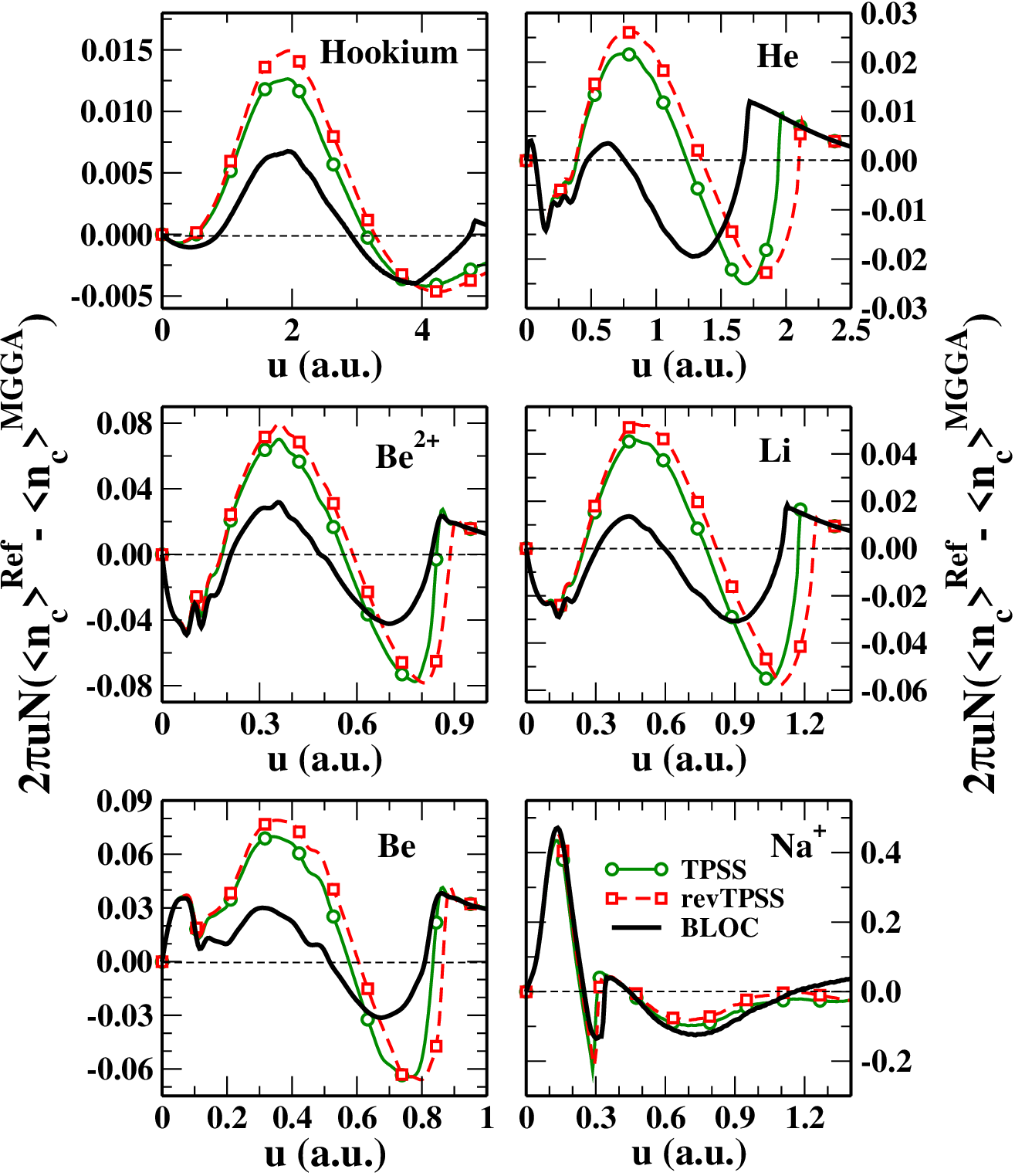}
\caption{\label{choles_fig}Errors in the real space analysis of the correlation energy of several 
atoms. Reference data are exact correlation holes from Refs. \onlinecite{hookium,hole3,hole6}, except 
for Na$^+$ in which case MP2 data were used \cite{hole6}.} 
\end{figure}

Figure \ref{choles_fig} reports the errors in the real-space analysis of the correlation energy of 
several atoms (Hookium, He, Be$^{2+}$, Li, Be, Na$^+$). 
Accurate reference correlation holes for these 
atoms were taken from Refs. \onlinecite{hookium,hole3,hole6}. 
Note that for Be and Na$^+$ the reference curve is
supposed to include also some static correlation that is not accounted by the dynamical correlation 
semilocal hole models. Thus, in these cases the comparison can be only qualitative. 

The plots for two- and three- electron systems show similar features,
displaying for all the
meta-GGAs negative errors at small and larger values of $u$ and a positive peak at medium values.
This behavior is directly related to the fact that the reference hole is, for these systems, more
localized that the semilocal ones: at small-$u$ values it is more negative than semilocal holes,
while at medium-$u$ values it is less negative than the meta-GGA correlation hole models.
The localization problem is of course more evident for TPSS and revTPSS, while it is less relevant
for BLOC, whose correlation was constructed with a proper localization constraint \cite{TPSSloc}.
In fact, the BLOC correlation hole is the most accurate for the considered systems.
%

\subsection{Exchange-correlation hole}
\label{xc_hole_sec}

%
\begin{figure}
\includegraphics[width=\columnwidth]{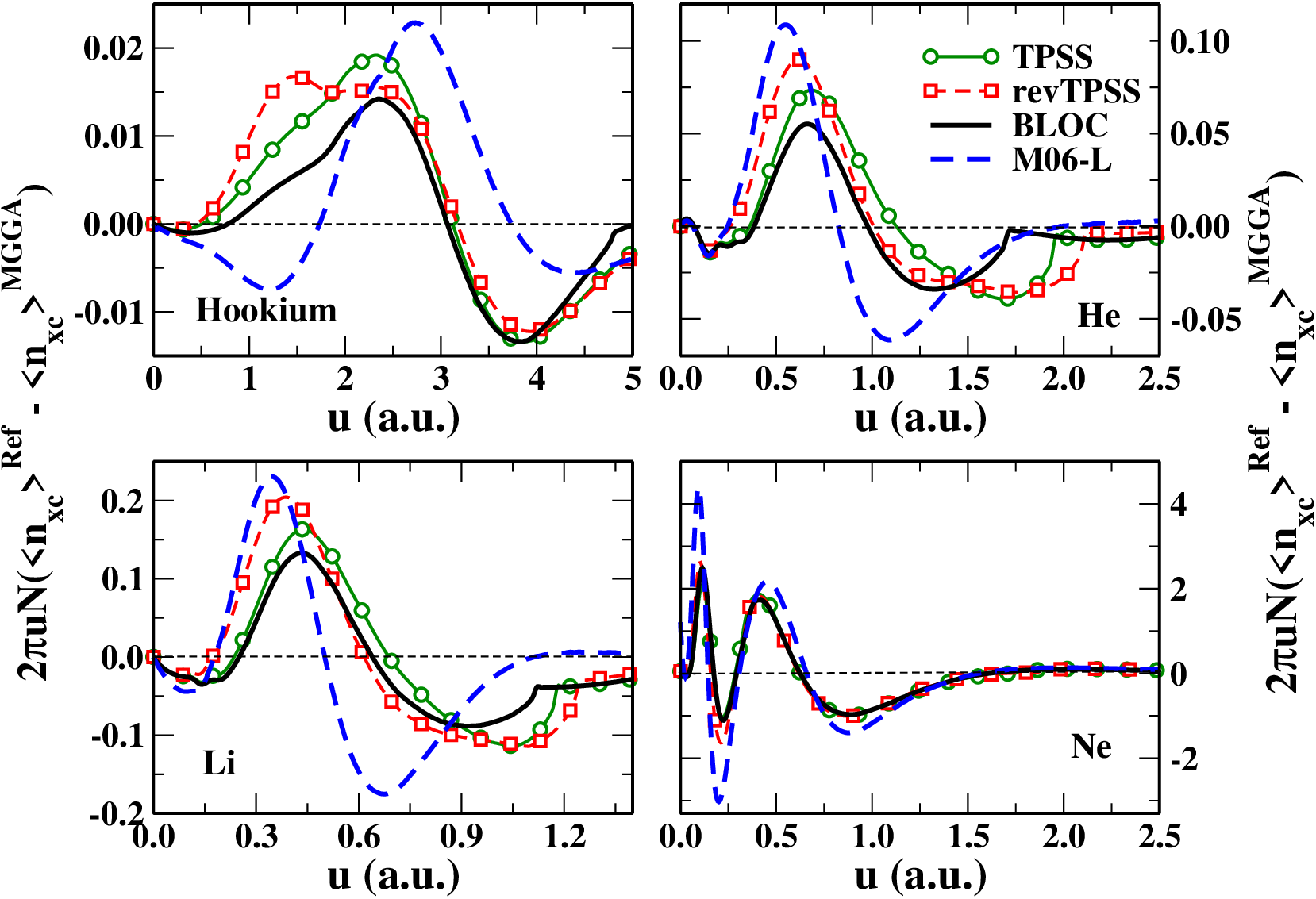}
\caption{\label{xc_fig} Errors in the real space analysis of the exchange-correlation energy of several
atoms.}
\end{figure}
%
In Fig. \ref{xc_fig} we report the errors in the real space analysis of
the XC energy of several atoms. 
The plots show that in all cases the BLOC meta-GGA   
gives overall the (slightly) best performance. However, in general, a similar
performance is found also for TPSS and revTPSS.
These results are in good agreement with those of Figs. \ref{xhole_atoms_fig}
and \ref{choles_fig}, where the separate behavior of the exchange and
correlation contributions was examined. Moreover, the plots of
Fig. \ref{xc_fig} indicate that the exchange and correlation holes of TPSS,
revTPSS and BLOC are well balanced with each other, and show a minimal accumulation
of errors between the exchange and correlation holes.

For further comparison we reported in Fig. \ref{xc_fig} also the errors
in the real-space analysis of the XC energy generated by the empirical M06-L 
meta-GGA functional \cite{m06-l}, which is widely used in DFT applications.
Because the M06-L exchange and correlation terms are not constructed
separately, only the full XC results are of interest. 
For simplicity, in applying the reverse-engineering method of Section
\ref{secii} to M06-L, we used the same function $\mathcal{H}(s,z)$ (see
Eq. (13)) as that used for the other non-empirical functionals (TPSS, revTPSS,
and BLOC). This simple expression in fact already provides an
accurate XC hole for M06-L, because of the recovering of 
energy and hole sum rules. 

The real-space analysis of the M06-L XC energy shows that this functional
yields a similar XC hole as TPSS and BLOC. Nevertheless, probably
because of its empirical nature, the M06-L functional displays a slightly
larger error compensation between the small and medium-large $u$ contributions.
Finally, it is interesting to note that, for real atoms, in comparison
with TPSS or BLOC, the errors in the real-space analysis of the M06-L XC
energy are much reduced at large $u$-values (the plots are ``compressed''
towards small values of $u$).
This feature may correlate with the fact that M06-L can describe better than
other functionals long-range effects (relevant for example for noncovalent and
dispersion interactions) \cite{marom11,zhao08_2}, which are dominant at large
$u$-values.
%

\section{Coulomb hole radius}
The correlation hole must fulfill the important constraint of the
particle sum rule, so that $\int \bar{n}_cd\mathbf{u}=0$. This condition 
implies that the correlation hole must be negative in some regions of space and positive in others.
More in detail, because the correlation on-top hole is well described by the LDA limit,
which is negative \cite{BEP}, we have that the
correlation hole will be negative for small $u$, will be equal to zero at some
distance, and then it will become positive for larger values of the electron-electron separation
(see Fig. \ref{comp_fig}). 
The position where the correlation hole equals zero for the first time
is defined as the Coulomb hole radius $R_c$.
An alternative definition can be given in terms of the $u$-distance where the XC and exchange-only
real-space analyses intersect for the first time.  

The physical meaning of the Coulomb hole radius is to define a sphere of radius $R_c$ around
an electron where the probability of finding another electron is reduced by the correlation
(because the correlation hole is negative inside this region). Analogously, outside the region
enclosed by the Coulomb hole radius ($u\geq R_c$) the probability to find a second electron 
is enhanced by the correlation effects \cite{hole1}.
The Coulomb hole radius is thus a meaningful quantity to be investigated.
However, it has not been given much attention at the DFT level.
Here we use the semilocal hole models to study this issue.

Table \ref{tab_rc} reports the Coulomb hole radii of several atoms and ions as computed
with different semilocal correlation functionals.
\begin{table}[bt]
\begin{center}
\caption{\label{tab_rc}Coulomb hole radius ($R_c$) as obtained from different semilocal 
approximations, for several atoms and ions. The last lines report the mean absolute error (MAE) and 
the mean absolute relative error (MARE). The results closest to the reference are denoted in bold 
style.}
\begin{ruledtabular}
\begin{tabular}{lrrrrrr}
 & LDA & PBE & TPSS & revTPSS & BLOC & Ref. \\ 
\hline
 \multicolumn{7}{c}{He isoelectronic series}\\
He & 3.51 & 1.32   & 1.17 & 1.23 & \bf{0.97} & 0.93$^a$ \\
Li$^{+}$ & 2.53 & 0.84  & 0.72 & 0.76 & \bf{0.64} & 0.67$^b$ \\
Be$^{+2}$ & 2.06 & 0.62  & 0.53 & 0.55 & \bf{0.48} & 0.49$^b$ \\
B$^{+3}$ & 1.77 & 0.49  & 0.42 & 0.43 & \bf{0.39} & 0.35$^c$ \\
Ne$^{+8}$ & 1.15 & 0.25  & \bf{0.20} & 0.21 & \bf{0.20} & 0.17$^c$ \\
Ca$^{+18}$ & 0.78 & 0.13  & \bf{0.10} & \bf{0.10} & \bf{0.10} & 0.08$^c$ \\ 
\hline
Hookium & 4.61 & 2.85  & 2.6 & 2.7 & \bf{2.4} & 2.25$^d$ \\
Li & 2.53 & 0.85  & 0.73 & 0.76 & \bf{0.64} & 0.67$^b$ \\
Be & 5.23 & 0.65  & 0.55 & 0.57 & \bf{0.50} & 0.48$^b$ \\
Ne & 2.35 & \bf{1.21}  & 1.10 & 1.14 & 1.05 & 1.27$^b$ \\
Na$^+$ & 2.05 & 1.10  & 0.94 & \bf{0.97} & 0.92 & 0.98$^b$ \\  \hline
MAE  & 1.84 & 0.18 & 0.10 & 0.12 & \bf{0.06} & \\
MARE & 3.91 & 0.30 & 0.15 & 0.18 & \bf{0.09} & \\
\end{tabular}
\end{ruledtabular}
\end{center}
\begin{flushleft}
a) Ref. \onlinecite{hole3}, b) Ref. \onlinecite{hole6}, c) Ref. \onlinecite{hole4}, d) Ref. 
\onlinecite{hookium}.
\end{flushleft}
\end{table}
The results show that all the functionals yield an overestimation for the
Coulomb hole radius in most of the systems. This behavior is related to the fact that
these functionals describe an insufficient localization for the correlation interaction,
tending to represent the correlation interaction as ``smeared out'' in space.
As a consequence, LDA fails badly, giving a mean 
absolute relative error (MARE) of about 400\%, but also the PBE correlation hole model \cite{PBW} 
overestimates $R_c$ by 30\% on average. Better results are obtained at meta-GGA level.
In particular, the BLOC correlation, which includes a localization constraint \cite{TPSSloc},
can describe the Coulomb hole radius with good accuracy.

This latter observation, applied to Be and Na$^+$ in particular, suggests that $R_c$ is a quantity 
mainly governed by the dynamical correlation, so that it can be well described by an accurate semilocal
dynamical correlation functional. Thus, static correlation may be supposed to act only on the depth 
and shape of the 
correlation hole inside the distance $R_c$.

\begin{figure}
\includegraphics[width=\columnwidth]{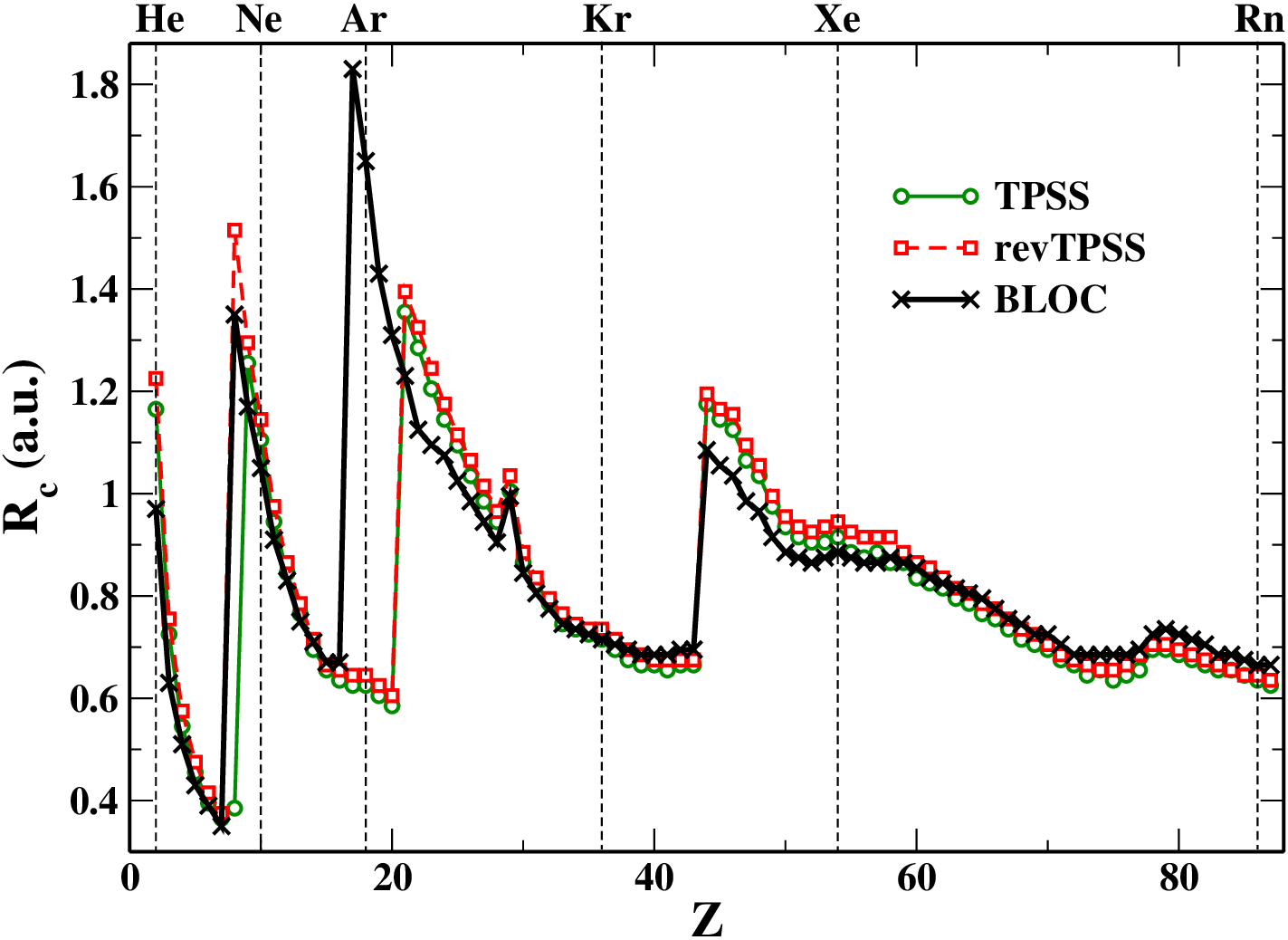}
\caption{\label{rc_atoms_fig}
Coulomb hole radius computed at the TPSS, revTPSS, and BLOC level for 
the atoms of the periodic table until $Z=87$.} 
\end{figure}

\begin{figure}
\includegraphics[width=\columnwidth]{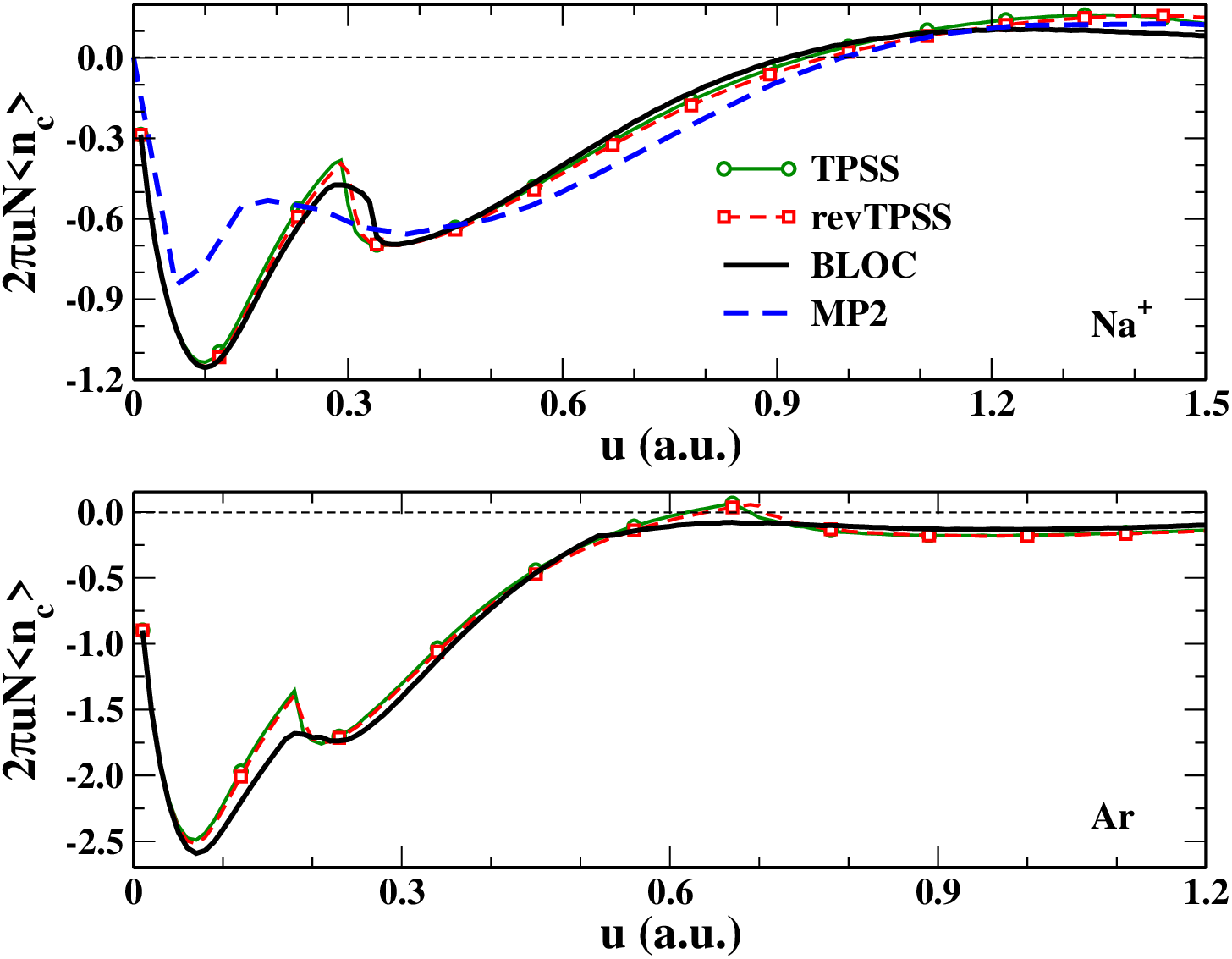}
\caption{\label{rc_compare_fig}
Real-space analysis of the correlation energy, for Na$^+$ (top) and Ar (down) atoms.} 
%
\end{figure}

To push the investigation further we computed the value of $R_c$ for the atoms of the periodic table
until Fr ($Z=87$), using TPSS, revTPSS, and BLOC (see Fig. \ref{rc_atoms_fig}).
We use KS exact exchange densities and orbitals.  
In this case no exact reference is available, thus the quality of the different results cannot
be assessed quantitatively. Nevertheless, it is possible to draw some interesting qualitative conclusions.
The data of Fig. \ref{rc_atoms_fig} show in fact that $R_c$ oscillates strongly for the lighter atoms
until $Z\approx 30$ (in agreement with the reference data reported in Table \ref{tab_rc}),
while after that the amplitude of the oscillations is 
dumped. The first peak in the magnitude of $R_c$ is at He ($Z=2$), the second one is at the O 
atom ($Z=8$) for BLOC and revTPSS and at the F atom ($Z=9$) for TPSS; the third peak is 
at Cl ($Z=17$) for BLOC and at the Sc atom ($Z=21$) for both TPSS and revTPSS.
Thus, interestingly for none of the functionals the trend follows exactly the periodicity 
of the periodic table. Moreover, we see that, although the three functionals display an overall 
similar trend,
important differences can be present for some atoms (for the O atom the value of $R_c$ differs
by more than 1.1 a.u. when computed at the TPSS and revTPSS level; a similar difference
is found for the $R_c$ of Cl computed at the BLOC and TPSS levels).
These features depend on the fact that for some ``critical'' atoms the value of $R_c$ is
determined by an inter-shell peak in the correlation hole that is below zero for some functionals 
and above it for others (see Fig. \ref{rc_compare_fig}).
%
Without additional benchmark calculations it is not possible to understand whose functional
is exhibiting the most correct behavior.

\begin{figure}
\includegraphics[width=\columnwidth]{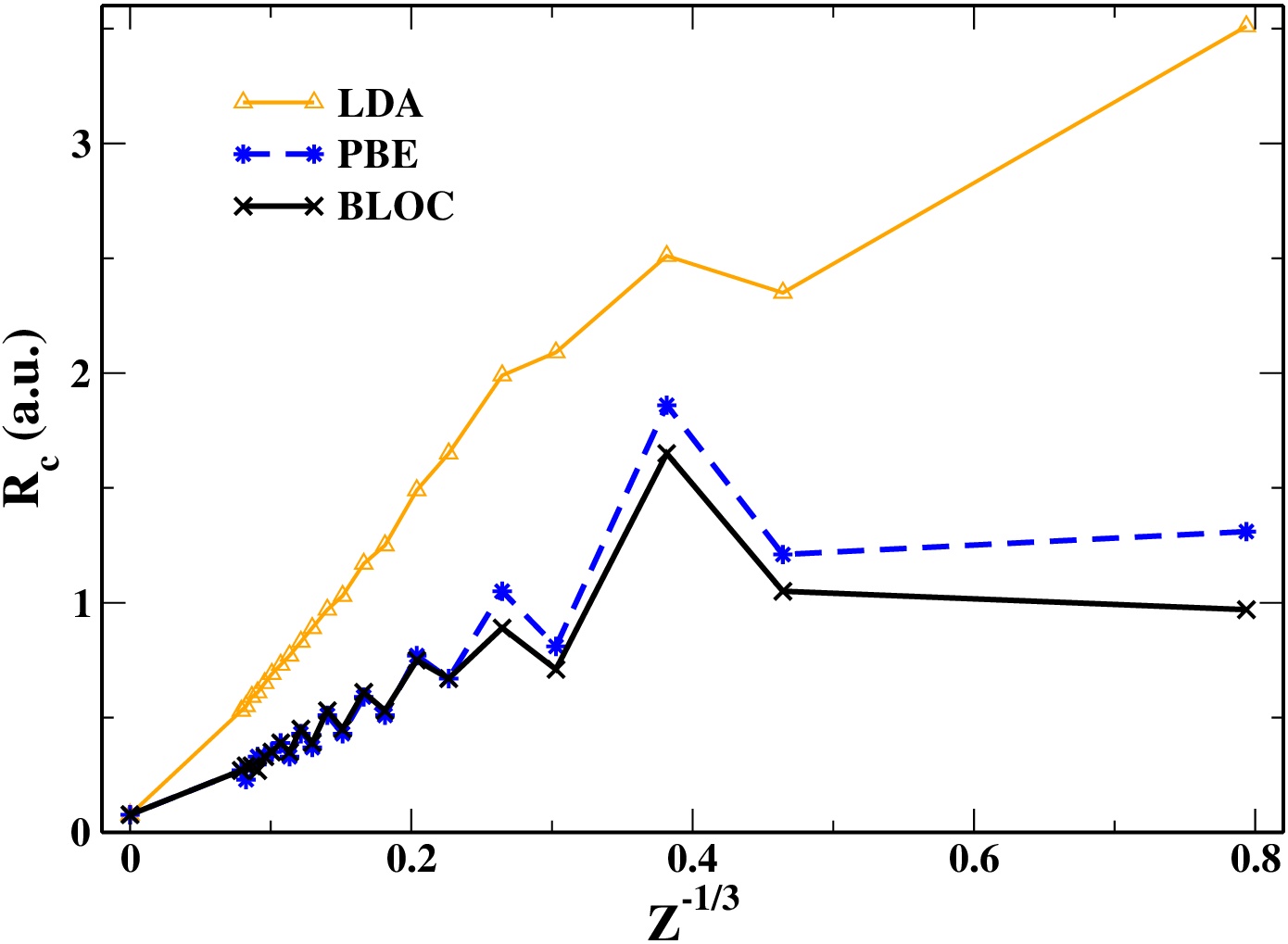}
\caption{\label{nob_rc_fig}Coulomb hole radius computed at the LDA, PBE, and BLOC level for noble gas 
atoms with $Z\leq 2022$. A parabolic extrapolation to $Z=\infty$ is also shown.} 
\end{figure}

To conclude our short investigation of the Coulomb hole radius we consider the problem
of how well local and semilocal approximations can perform for atoms with an increasing number of electrons.
To this end we report in Fig. \ref{nob_rc_fig} the values of $R_c$ computed at the LDA, PBE, and
BLOC level, for noble atoms with up to 2022 electrons.
The calculations were performed using exact exchange orbitals and densities, similarly with the ones 
of Refs. \cite{ioni,APBEK}. 
Inspection of the figure shows that LDA gives a very diffuse correlation hole for all 
the atoms (see also Table \ref{tab_rc}), and only in the limit $Z=\infty$ may become exact 
\cite{ELCB,scal}.
On the contrary, PBE yields a visible overestimation with respect to BLOC for most atoms in the
periodic table, but the two methods agree well for atoms with $Z\gtrapprox 100$.  
This suggests that in the limit of large atoms the meta-GGA and GGA levels may
be equivalent for the description of the Coulomb hole radius.
Finally, we note that for heavy atoms $R_c$ is small in $u$-space, but not in $v$-space.
In fact the LDA hole \cite{PW} in the high-density limit has $R_c$ proportional with the 
Thomas-Fermi wave vector $k_s$.

\section{Conclusions}
In this paper we have developed a generally-applicable meta-GGA XC hole model 
based on a reverse-engineering  method. It recovers as many exact constraints as a 
meta-GGA hole can and it is very realistic at small and medium electron-electron distances. 
However, further work can be forseen to improve the large-$u$ behavior
of the hole model. 
In fact, the behavior of a semilocal hole at large electron-electron displacements
can be hardly described with accuracy because it is dominated by highly nonlocal XC effects.
This is evident, for example, in the dissociation limit of the hydrogen molecule
where the exact XC hole is strongly asymmetrical \cite{vonBarth1} in contrast to the
semilocal hole which is by construction spherically symmetric.
Further difficulties might be encountered by the present semilocal models in the description of 
low-dimensional problems (e.g. quasi-two-dimensional regime), where additional constraints
need to be satisfied \cite{q2D}.

Despite the limitations mentioned above the hole model developed here is anyway a
powerful tool for DFT. In fact, we have applied it to the TPSS, revTPSS and BLOC meta-GGA XC functionals, 
and we  have shown that the resulting meta-GGA isotropic Coulomb hole is realistic in many 
situations, giving accurate XC real-space analysis of small atoms and ions  as well as XC wave-vector 
analysis of jellium surfaces. Similar results were also demonstrated for the
M06-L XC hole.
Moreover, we considered the evaluation of the Coulomb hole radius for various atomic
systems, showing that this quantity, which was investigated until now only by high-level, 
wave function methods, being almost out of reach for DFT, has a great interest for the
detailed understanding of the correlation interaction and the development of better approximations. 
Also in this case the meta-GGA correlation hole models, in particular the BLOC one, proved to be
efficient tools for the calculation of the Coulomb hole radius, yielding errors less
than 10\% with respect to benchmark data.

Finally, we recall that the present hole model can be a good starting point for 
future developments beyond the semilocal level. It can be used in fact in the development of
higher-level non-local orbital-dependent methods, satisfying exact conditions
beyond those available to a meta-GGA functional. In this sense, screened hybrid functionals \cite{krukau08}
are the most popular example. However, recently also more advanced models have been proposed
such as the screened exchange model of Ref. \onlinecite{vonBarth1}, that reads
\begin{equation}
E_{xc} = -\frac{1}{4}\int\frac{\gamma(\R,\R')}{|\R-\R'|}F(\R,\R')d\R d\R'\ ,
\end{equation}
where $\gamma(\R,\R')$ is the spin-resolved non-interacting density matrix, and
$F(\R,\R')\equiv \bar{n}_{xc}(\R,\R')/\bar{n}_{x}(\R,\R')$ is a
screening factor that may be efficiently modeled by considering accurate semilocal 
exchange and correlation hole models.

\begin{acknowledgment}
This work was funded by the ERC Starting Grant FP7
Project DEDOM, Grant Agreement No. 207441.
\end{acknowledgment}

\end{document}